# Distributing Content Simplifies ISP Traffic Engineering

Abhigyan Sharma    Arun Venkataramani    Ramesh Sitaraman

University of Massachusetts Amherst
{abhigyan, arun, ramesh}@cs.umass.edu

## ABSTRACT

Several major Internet service providers (e.g., Level-3, AT&T, Verizon) today also offer content distribution services. The emergence of such "Network-CDNs" (NCDNs) are driven by market forces that place more value on content services than just carrying the bits. NCDNs are also necessitated by the need to reduce the cost of carrying ever-increasing volumes of traffic across their backbones. An NCDN has the flexibility to determine both where content is placed and how traffic is routed within the network. However NCDNs today continue to treat traffic engineering independently from content placement and request redirection decisions. In this paper, we investigate the interplay between content distribution strategies and traffic engineering and ask how an NCDN should engineer traffic in a content-aware manner. Our experimental analysis, based on traces from a large content distribution network and real ISP topologies, shows that effective content placement can significantly simplify traffic engineering and in most cases obviate the need to engineer NCDN traffic all together! Further, we show that simple demand-oblivious schemes for routing and placement such as InverseCap and LRU suffice as they achieve network costs that are close to the best possible.

## 1. INTRODUCTION

Content delivery networks (CDNs) today provide a core service that enterprises use to deliver web content, downloads, streaming media, and IP-based applications to a global audience of their end-users. The traditional and somewhat simplified, tripartite view of content delivery involves three sets of entities as shown in Figure 1. The *content providers* (e.g., media companies, news channels, e-commerce providers, software distributors, enterprise portals, etc.) produce the content and wish to provide a high-quality experience to end-users accessing their content over the Internet. The *networks* (e.g., telcos such as AT&T, MSOs such as Comcast, and traditional ISPs) own the underlying network infrastructure and are responsible for provisioning capacity and managing traffic flowing through their networks. Finally, the *CDNs* (e.g., Akamai, Limelight) are responsible for optimizing content delivery to end-users on behalf of the content providers, residing as a global, distributed overlay service on top of the networks.

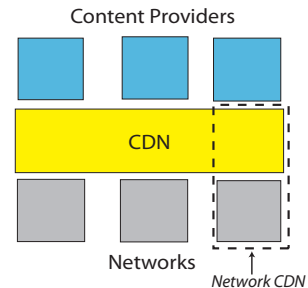

Figure 1: **A tripartite view of content delivery.**

Recent powerful trends are reshaping the simplified tripartite view of content delivery. A primary driver is the torrid growth of video [22, 8] and downloads traffic on the Internet. For example, a single, popular TV show with 50 million viewers, each viewer watching an HD-quality stream of 10 Mbps, generates 500 Tbps of network traffic! The increasing migration of traditional media content to the Internet and the consequent challenges of scaling the network backbone to accommodate that traffic has necessitated the evolution of *network CDNs* (or NCDNs)[1] that vertically integrate CDN functionality such as content caching and redirection with traditional network operations [15, 21, 20, 5, 28] (refer Figure 1). A second economic driver of NCDNs is the desire of networks to further monetize the "bits" that flow on their infrastructure by contracting directly with content providers. Finally, NCDNs also enable better performance for their own end-user subscribers and open up avenues for new, differentiated services (e.g., Verizon's recent offering that delivers HBO's content to FIOS subscribers [27].

The evolution of NCDNs significantly changes traditional engineering concerns as NCDNs must make decisions about content placement and request redirection in addition to the underlying network routing. As NCDNs own both the content distribution and network infrastructure, they are in a powerful position to place content in a manner that "shapes" the traffic demand to their advantage, potentially enabling traffic engineering to achieve a significantly lower cost. However, NCDNs today treat content distribution and traffic engineering concerns separately, perhaps because it is easier to continue doing things as they were being done. This disparity raises several research questions such as: (1) How do content demand patterns and placement strategies impact traffic engineering objectives? (2) How should an NCDN jointly determine placement and routing decisions so as optimize network cost? (3) How do demand-aware strategies (i.e., using knowledge of recently observed demand patterns or hints about anticipated future demands) for placement and routing compare with demand-oblivious strategies?

---

[1]NCDNs are sometimes referred to as Telco CDNs, or Carrier CDNs. Further, they are referred to as a Licensed CDN when a pure-play CDN such as Edgecast[7] licenses the CDN software to a network to create an NCDN.

Our primary contribution is to empirically analyze the above questions for realistic content demand workloads and ISP topologies. To this end, we collect content request traces from Akamai, the world's largest CDN today. We focus specifically on on-demand video and large-file downloads traffic as they are two categories that dominate overall CDN traffic and are significantly influenced by content placement strategies. Our combined traces consist of a total of 28.2 million requests from 7.79 million unique users who downloaded a total of 1455 Terabytes of content across the US over multiple days. Our trace-driven experiments using these logs and realistic ISP topologies reveal the following somewhat surprising conclusions:

- Simple demand-oblivious schemes for placement and routing (such as Least Recently Used and Inverse-Cap) significantly outperform (by 2.2× to 17×) a joint-optimal placement and routing with knowledge of the previous day's demand[2].
- Traffic demand can be "shaped" by effective content placement so that traffic engineering, i.e., optimizing routes with knowledge of recent traffic matrices, yields little improvement in cost compared to demand-oblivious routing (InverseCap) in conjunction with any reasonable placement.
- A demand-oblivious placement and routing is at most 4% sub-optimal compared to a joint-optimal placement and routing with perfect knowledge of the next day's demand at higher storage ratios ($\approx 4$) with simple optimizations such as content chunking and link-utilization-aware redirection.

The paper outline is as follows. First, (§2) provides an overview of the NCDN architecture. Next, we formulate algorithms that jointly optimize content placement and routing in an NCDN (§3). We then describe how we collected real CDN traces (§4) and then evaluate our algorithms using these traces and real ISP topologies (§5). Finally, we present related work (§6) and conclusions (§7).

---
[2]We use the term "optimal" when placement or routing is the solution of an optimization problem, but the solution may not have the lowest cost (for reasons detailed in §5.3)

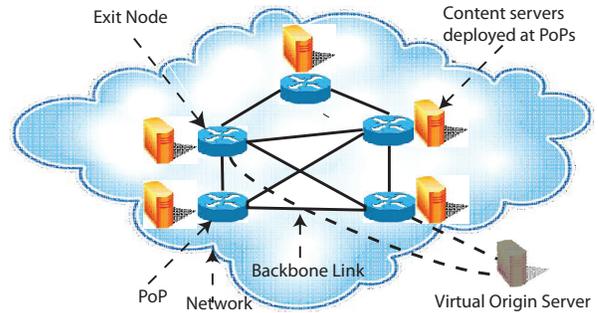

Figure 2: NCDN Architecture

## 2. BACKGROUND AND MOTIVATION

A typical NCDN architecture, as shown in Figure 2, resembles the architecture of a global CDN but with some important differences. First, the content servers are deployed at points-of-presence (PoPs) within the network rather than globally across the Internet as the NCDN is primarily interested in optimizing content delivery for its own customers and end-users. Second, and more importantly, the NCDN owns and manages the content servers as well as the underlying network. Content providers that purchase content delivery service from the NCDN publish their content to origin servers that they maintain external to the NCDN itself.

Each PoP is associated with a distinct set of end-users who request content such as web, video, downloads etc. An end-user's request is first routed to the content servers at the PoP to which the end-user is connected. If a content server at that PoP has the requested content in their cache, it serves that to the end-user. Otherwise, if the requested content is cached at other PoPs, the content is downloaded from a nearby PoP and served to the end-user. If the content is not cached in any PoP, it is downloaded directly from the content provider's origin servers.

### 2.1 Why Do NCDNs Change the Game?

Managing content distribution as well as the underlying network makes the costs and objectives of interest to an NCDN different from that of a traditional CDN or a traditional ISP. The traditional model of content distribution and traffic engineering as performed by a traditional CDN and a traditional ISP is shown in Figure 3. We propose a new model appropriate for NCDNs that jointly optimizes content distribution and traffic engineering as shown in Figure 4.

#### 2.1.1 Content Distribution

A traditional CDN has two key decision components—

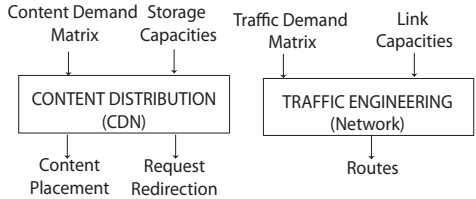

Figure 3: Traditional formulation with content distribution and traffic engineering optimized separately.

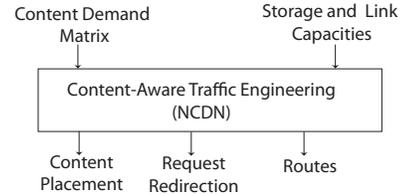

Figure 4: Our new formulation of content-aware traffic engineering for NCDNs as a joint optimization.

*content placement* and *request redirection*—that seek to optimize the response time perceived by end-users and balance the load across its content servers. Content placement decides which objects should be cached at which nodes. An object may be stored at multiple nodes in the network or not stored in the network at all and be served from the origin server instead. Request redirection determines which server storing a replica of the requested object is best positioned to serve the request.

Content placement schemes are either *demand-aware* or *demand-oblivious*. A demand-aware scheme calculates placement using a *content matrix* that specifies the demand for each content at each content server location. The content matrix is learned by monitoring a recent history of system-wide requests and possibly including hints, if any, from content providers about anticipated demand for some objects. A demand-aware scheme uses a recent content matrix to decide on a placement periodically (say, once a day) but does not alter its placement in between. In contrast, a demand-oblivious scheme can continually alter its placement potentially even after every single request. A simple example of a demand-oblivious placement scheme is least-recently-used (LRU), where each node manages its cache using LRU cache replacement strategy.

### 2.1.2 Traffic Engineering

A key component of ISP network operations is traffic engineering, which seeks to route the traffic demands through the backbone network so as to balance the load and mitigate hotspots. Traffic engineering is commonly viewed as an routing problem that takes as input a *traffic matrix*, i.e., the aggregate flow demand between every pair of PoPs observed over a recent history, and computes routes so as to minimize a network-wide cost objective. The cost seeks to capture the severity of load imbalance in the network and common objective functions include the maximum link utilization (MLU) or a convex function (so as to penalize higher utilization more) of the link utilization aggregated across all links in the network [12]. ISPs commonly achieve the computed routing either by using shortest-path routing (e.g., the widely deployed OSPF protocol [12]) or by explicitly establishing virtual circuits (e.g., using MPLS [11]). ISPs perform traffic engineering at most a few times each day, e.g., morning and evening each day [13].

Routing can also be classified as *demand-aware* or *demand-oblivious* similar in spirit to content placement. Traffic engineering schemes as explained above (including online traffic engineering schemes [19] that are rarely deployed) are implicitly demand-aware as they optimize routing for recently observed demand. In contrast, demand-oblivious routing schemes rely upon statically configured routes [4, 6], e.g., InverseCap does static shortest-path routing in which link weights are set to inverse of link capacities. This is the default weight setting for OSPF in Cisco routers [13].

### 2.1.3 Content-aware Traffic Engineering

An NCDN can perform content-aware traffic engineering by leveraging content distribution to achieve traffic engineering goals such as network cost minimization. Unlike traditional ISPs, an NCDN can place content and redirect requests in a manner that "shapes" the traffic demands to its advantage and thereby achieve significantly lower network cost.

A central thesis of this paper is that intelligent content placement and request redirection achieve significant cost reduction for NCDNs. To appreciate this point, consider the simple, illustrative example in Figure 5. Node $C$ has an object in its cache that is requested by end-users at nodes $A$ and $D$. Suppose that one unit of traffic needs to be routed from $C$ to $A$ and 0.5 units from $C$ to $D$ to satisfy the demand for that object. The routing that achieves the minimum MLU of 0.5 to serve the demanded object is shown in the figure. Note that the routing that achieves the MLU of 0.5 is not possible with a simple, demand-oblivious protocol like InverseCap as that would route all the traffic demand from $C$ to $A$ via $B$, resulting in an MLU of 1. Thus, a (demand-aware) traffic engineering scheme is necessary to achieve an MLU of 0.5.

On the other hand, content-aware traffic engineering can shape the traffic demand matrix by using a judicious placement and redirection strategy. Suppose that there is some space left in the content server's cache at node $B$ to accommodate an additional copy of the demanded object. By creating an additional copy of the object at $B$, the traffic demand of $A$ can be satisfied from $B$ and the demand of $D$ from $C$ achieving the an MLU of 0.125. In this case, judicious content placement decreased the MLU by a factor of 4. Even more interestingly, this best MLU can be achieved using a simple routing scheme like InverseCap. Although this is clearly a "toy" example, it illustrates the sophisticated interaction between content placement and traffic engineering and potential opportunities for NCDNs to both reduce cost and simplify traffic engineering by doing it in a content-aware manner.

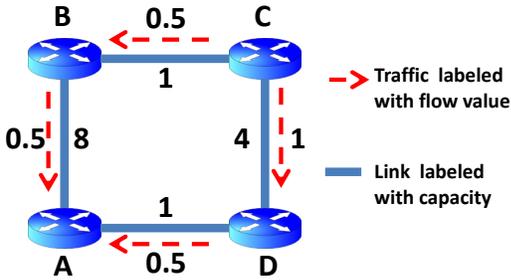

Figure 5: A simple NCDN example

| Input variables and descriptions | |
|---|---|
| $V$ | Set of nodes where each node represents a PoP |
| $E$ | Set of edges where each link represents a communication link |
| $o$ | Virtual origin node that hosts all the content in $K$ |
| $X$ | Set of exit nodes in $V$ |
| $D_i$ | Disk capacity at node $i \in V$ (in bytes) |
| $C_e$ | Capacity of link $e \in E$ (in bits/sec) |
| $K$ | the set of all content accessed by end-users |
| $S_k$ | Size of content $k \in K$. |
| $T_{ik}$ | Demand (in bits/sec) at node $i \in V$ for content $k \in K$ |
| **Decision variables and descriptions** | |
| $\alpha$ | MLU of the network |
| $z_k$ | Binary variable indicating whether one or more copies of content $k$ is placed in the network |
| $x_{jk}$ | Binary variable indicating whether content $k$ is placed at node $j \in V \cup \{o\}$ |
| $f_{ij}$ | Total traffic from node $j$ to node $i$ |
| $f_{ije}$ | Traffic from node $j$ to node $i$ crossing link $e$. |
| $t_{ijk}$ | Traffic demand at node $i \in V$ for content $k \in K$ served from node $j \in V \cup \{o\}$ |

Table 1: List of input and decision variables for the NCDN problem formulation.

## 3. CONTENT-AWARE TRAFFIC ENGINEERING FOR AN NCDN

In this section, we formalize the NCDN model and the problem of determining the optimal content placement and routing as a mixed integer program (MIP). Next, we present hardness and inapproximability results for this problem. Finally, we discuss approximation techniques that we use to solve our MIP formulation.

### 3.1 NCDN Model

Table 1 lists all the model parameters. An NCDN consists of a set of nodes $V$ where each node represents a PoP in the network. The nodes are connected by a set of directed edges $E$ that represent the backbone links in the network. The set of content requested by end-users is represented by the set $K$ and the sizes of content are denoted by $S_k, k \in K$. The primary resource constraints are the link capacities $C_e, e \in E$, and the storage at the nodes $D_i, i \in V$. We implicitly assume that the content servers at the PoPs have adequate compute resources to serve locally stored content.

A *content matrix* specifies the demand for each content at each node. An entry in this matrix, $T_{ik}, i \in V, k \in K$, denotes the demand (in bits/second) for content $k$ at node $i$. The content matrix is assumed to be measured by the NCDN a priori over a coarse-grained interval, e.g., the previous day. The infrastructure required for this measurement is comparable to what ISPs have in place for monitoring traffic matrices today.

Origin servers, owned and maintained by the NCDN's content providers, initially store all content published by content providers. We model origin servers using a single virtual origin node $o$ external to the NCDN that can be reached via a set of exit nodes $X \subset V$ in the NCDN (see Figure 2). Since we are not concerned with traffic engineering links outside the NCDN, we model the edges $(x, o)$, for all $x \in X$, as having infinite capacity. The virtual origin node $o$ always maintains a copy of all the requested content. However, a request for a content is served from the virtual origin node only if no copy of the content is stored at any node $i \in V$. In this case, the request is assumed to be routed to the virtual origin via the exit node closest to the node where the request was made (in keeping with the commonly practiced *early-exit* or *hot potato* routing policy).

### 3.2 Optimal Strategy as MIP

The problem of content-aware traffic engineering by NCDNs (*NCDN problem* for short) seeks to compute a placement and routing that minimizes the MLU and satisfies the demands specified by the content matrix while respecting link capacity and storage constraints. This optimization goal can be formulated as a mixed integer program (MIP). Unlike the traditional traffic engineering problem that can be formulated as a multi-commodity flow problem and solved using a linear program, the NCDN problem needs to make binary placement decisions, i.e., whether or not to place a content at a PoP, and then route the demand accordingly. The placement decision variables and other decision variables for this problem are listed in Table 1. The MIP to minimize the MLU $\alpha$ is as follows:

$$\min \alpha \quad (1)$$

subject to

$$\sum_{j \in V} t_{ijk} + t_{iok} = T_{ik}, \quad \forall k \in K, i \in V \quad (2)$$

$$\sum_{k \in K} t_{ijk} = f_{ij}, \quad \forall j \in V - X, i \in V \quad (3)$$

$$\sum_{k \in K} t_{ijk} + \sum_{k \in K} \delta_{ij} t_{iok} = f_{ij}, \quad \forall j \in X, i \in V \quad (4)$$

where $\delta_{ij}$ is 1 if $j$ is the closest exit node to $i$ and 0 otherwise. Note that $\delta_{ij}$ is not a variable but a constant that is determined by the topology of the network, and hence constraint (4) is linear.

$$\sum_{p \in P(l)} f_{ijp} - \sum_{q \in Q(l)} f_{ijq} = \begin{cases} f_{ij} & \text{if } l = i, \\ -f_{ij} & \text{if } l = j, \\ 0 & \text{otherwise,} \end{cases} \quad \forall i, j, l \in V \quad (5)$$

where $P(l)$ and $Q(l)$ respectively denote the set of outgoing and incoming links at node $l$.

$$\sum_{i \in V, j \in V} f_{ije} \leq \alpha \times C_e, \quad \forall e \in E \quad (6)$$

$$\sum_{k \in K} x_{ik} S_k \leq D_i, \quad \forall i \in V \quad (7)$$

$$x_{ok} = 1, \quad \forall k \in K \quad (8)$$

$$\sum_{i \in V} x_{ik} \geq z_k, \quad \forall k \in K \quad (9)$$

$$x_{ik} \leq z_k, \quad \forall k \in K, i \in V \quad (10)$$

$$t_{ijk} \leq x_{jk} T_{ik}, \quad \forall k \in K, i \in V, j \in V \cup \{o\} \quad (11)$$

$$t_{iok} \leq T_{ik}(1 - z_k), \quad \forall k \in K \quad (12)$$

$$x_{jk}, z_k \in \{0, 1\}, \quad \forall j \in V, k \in K$$

$$f_{ije}, t_{ijk}, t_{iok} \geq 0, \quad \forall i, j \in V, e \in E, k \in K$$

The constraints have the following rationale. Constraint (2) specifies that the total traffic demand at each node for each content must be satisfied. Constraints (3) and (4) specify that the total traffic from source $j$ to sink $i$ is the sum over all content $k$ of the traffic from $j$ to $i$ for $k$. Constraint (5) specifies that the volume of a flow coming in must equal that going out at each node other than the source or the sink. Constraint (6) specifies that the total flow on a link is at most $\alpha$ times capacity. Constraint (7) specifies that the total size of all content stored at a node must be less than its disk capacity. Constraint (8) specifies that all content is placed at the virtual origin node $o$. Constraints (9) and (10) specify that at least one copy of content $k$ is placed within the network if $z_k = 1$, otherwise $z_k = 0$ and no copies of $k$ are placed at any node. Constraint (11) specifies that the flow from a source to a sink for some content should be zero if the content is not placed at the source (i.e., when $x_{jk} = 0$), and the flow should be at most the demand if the content is placed at the source (i.e., when $x_{jk} = 1$). Constraint (12) specifies that if some content is placed within the network, the traffic from the origin for that content must be zero. Updating the content placement itself generate traffic and impacts the MLU in the network. We have omitted a formal description of the corresponding constraints for ease of exposition and lack of space.

### 3.3 Computational Hardness

Opt-NCDN is the decision version of the NCDN problem. The proofs for these theorems are presented in [1].

THEOREM 1. *Opt-NCDN is NP-Complete even in the special case where all objects have unit size, all demands have binary values, and link and storage capacities have binary values.*

COROLLARY 1. *Opt-NCDN is inapproximable to within a constant factor unless $P = NP$.*

### 3.4 Approximation Techniques for MIP

As solving the MIP for very large problem scenarios is computationally infeasible, we use two approximation techniques to tackle such scenarios.

The first is a two-step local search technique. In the first step, we "relax" the MIP by allowing the integral variables $x_{jk}$ and $z_k$ to take fractional values between 0 and 1. This converts an MIP into an LP that is more easily solvable. Note also that the optimal solution of the relaxed LP is a lower bound on the optimal solution of the MIP. However, the LP solution may contain fractional placement of some of the content with the corresponding $x_{jk}$ variables set to fractional values between 0 and 1. However, in our experiments only about 20% of the variables in the optimal LP solution were set to fractional values between 0 or 1, and the rest took integral values of 0 or 1. In the second step, we search for a valid solution for the MIP in the local vicinity of the LP solution by substituting the values for variables that were set to 0 or 1 in the LP solution, and re-solving the MIP for the remaining variables. Since the number of integer variables in the second MIP is much smaller, it can be solved more efficiently than the original MIP.

The second approximation technique is to reduce the number of unique content in the optimization problem using two strategies. First, we discard the tail of unpopular content prior to optimization. The discarded portion accounts for only 1% of all requests, but reduces the number of content by 50% or more in our traces. Second, we sample 25% of the content from the trace and, in our experiments, select trace entries corresponding only to the sampled content. These approximations reduce the number of content from tens of thousands to less than 5000. An MIP of this size can be solved using local search within an hour by a standard LP solver [16] for the ISP topologies in our experiments. To check for any untoward bias introduced by the sampling, we also performed a small number of experiments with the complete trace and verified that our conclusions remain qualitatively unchanged.

## 4. AKAMAI CDN TRACES

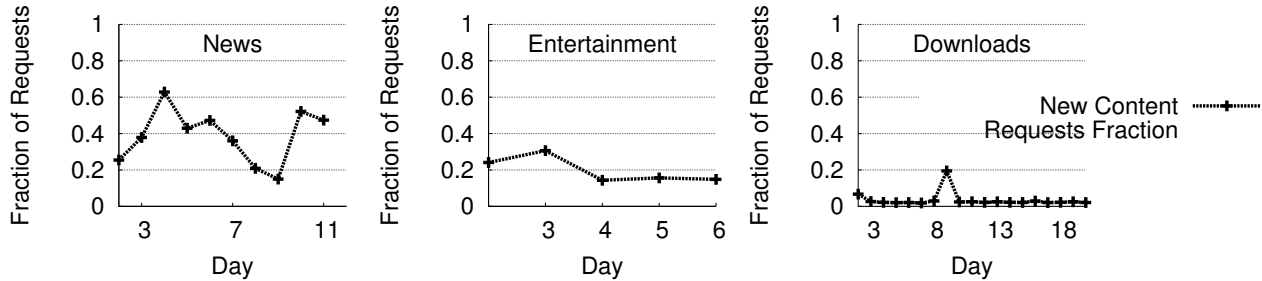

Figure 6: News and entertainment have a significant fraction of requests for new content on all days. Downloads has a small fraction of requests for new content on all days, except one.

To conduct a realistic simulation of end-users accessing content on an NCDN, we collected extensive traces of video and downloads traffic from Akamai, one of the world's largest CDNs.

**Video traces.** Videos are the primary source of traffic on a CDN and are growing at a rapid rate [22, 8]. Our video trace consists of actual end-users accessing on-demand videos on the Akamai network over multiple days. To make the traces as representative as possible, we chose content providers with a whole range of business models, including major television networks, news outlets, and movie portals. The videos in our traces include a range of video types from short-duration video (less than 10 mins) such as news clips to longer duration (30 min to 120 min) entertainment videos representing TV shows and movies. In all, our traces represent a nontrivial fraction of the overall traffic on Akamai's media network and accounted for a total of 27 million playbacks of over 85000 videos, 738 TBytes of traffic, served to 6.59 million unique end-users around the US. Since we only had US-based network topologies, we restricted ourselves to US-based traffic.

We collect two sets of video traces called *news trace* and *entertainment trace* respectively. The news trace was collected from a leading news outlet for an 11-day period in Sept 2011, and consists mostly of news video clips, but also includes a small fraction of news TV shows. The entertainment trace was collected for a 6 day period in January 2012, and includes a variety of videos including TV shows, clips of TV shows, movies and movie trailers from three major content providers.

The trace collection mechanism utilized a plugin embedded in the media player that is capable of reporting (anonymized) video playback information. Our traces include a single log entry for each video playback and provides time of access, user id, the location of the user (unique id, city, state, country, latitude, and longitude), the url of the content, the content provider, the total length of the video (in time and bytes), the number of bytes actually downloaded, the playback duration, and the average bitrate over the playback session.

**Downloads traces.** The second largest traffic contributor in a CDN is downloads of large files over HTTP. These include software and security updates, e.g., Microsoft's Windows or Symantec's security updates, as well as music, books, movies, etc.. The large file downloads at Akamai typically use a client-side software called the download manager [23]. We collect extensive and anonymized access data reported from the download manager using Akamai's NetSession interface [17] for a large fraction of content providers for a period of a month (December 2010). Our traces represent a nontrivial fraction of the overall US-based traffic on Akamai's downloads network and accounted for a total of 1.2 million downloads, 717 TBytes of traffic, served to 0.62 million unique end-users around the US. Our traces provide a single log entry for each download and provide time of access, user id, location of the user (city, state, country, latitude, and longitude), the url identifier of the content, content provider, bytes downloaded, and file size.

Figure 6 shows the fraction of requests for new content published each day relative to the previous day for news, entertainment, and downloads traces. The news trace has up to 63% requests due to new content because the latest news clips generated each day are the most popular videos on the website. The entertainment trace also has up to 31% requests each day due to new content such as new episodes of TV shows, and the previews of upcoming TV shows. The downloads trace has only 2-3% requests due to new content on a typical day. However, on the 9th day of the trace major software updates were released, which were downloaded on the same day by a large number of users. Hence, nearly 20% requests on that day were for new content. The fraction of requests for new content impacts the performance of demand-aware placement strategies as we show §5.

## 5. EXPERIMENTAL EVALUATION

We conduct trace-driven experiments to compare different content-aware traffic engineering strategies for NCDNs. Our high-level goal is to identify a simple strategy that performs well for a variety of workloads. In addition, we seek to assess the relative value of optimizing content placement versus routing; the value of being demand-aware versus being demand-oblivious; and the value of future knowledge about demand.

## 5.1 Trace-driven Experimental Methodology

To realistically simulate end-users accessing content on an NCDN, we combine the CDN traces (in §4) with ISP topologies as follows. We map each content request entry in the Akamai trace to the geographically closest PoP in the ISP topology in the experiment (irrespective of the real ISP that originated the request). Each PoP has a content server as shown in Figure 2, and the request is served locally, redirected to the nearest (by hop-count) PoP with a copy, or to the origin as needed.

**ISP topologies.** We experimented with network topology maps from two US-based ISPs. The first is the Abilene topology [26] and the second is a large tier-1 US ISP topology (referred to as US-ISP).

**MLU computation.** We compute the traffic that flow through each link periodically. To serve a requested piece of content from a PoP $s$ to $t$, we update the traffic induced along all edges on the path(s) from $s$ to $t$ as determined by the routing protocol using the bytes-downloaded information in the trace. To compute the MLU, we partition simulation time into 5-minute intervals and compute the average utilization of each link in each 5-minute interval. We discard the values of the first day of the trace in order to warm up the caches, as we are interested in steady-state behavior. We then compute our primary metric, which is the 99-percentile MLU, as the $99^{th}$ percentile of the link utilization over all links and all 5-minute time periods. We use 99-percentile instead of the maximum as the former is good proxy for the latter but with less experimental noise. Finally, for ease of visualization, we scale the 99-percentile MLU values in all graphs so that the maximum 99-percentile MLU across all schemes in each graph is equal to 1. We call this scaled MLU the *normalized MLU*. Note that only the relative ratios of the MLUs for the different schemes matter and scaling up the MLU uniformly across all schemes is equivalent to uniformly scaling down the network resources or uniformly scaling up the traffic in the CDN traces.

**Storage.** We assume that storage is provisioned uniformly across PoPs except in §5.7 where we analyze heterogenous storage distributions. We repeat each simulation with different levels of provisioned storage. Since the appropriate amount of storage depends on the size of the working set of the content being served, we use as a metric of storage the *storage ratio*, or the ratio of total storage at all PoPs in the network to the average storage footprint of all content accessed in a day for the trace. The total storage across all nodes for a storage ratio of 1 is 228 GB, 250 GB, and 895 GB for news, entertainment and downloads respectively.

## 5.2 Schemes Evaluated

Each evaluated scheme has a content placement component and a routing component.

**InvCap-LRU** uses LRU as the cache replacement strategy and InverseCap (with ECMP) as the routing strategy. InverseCap is a static, shortest-path routing scheme where link weights are set to the inverse of the link capacity. This scheme requires no information of either the content demand or the traffic matrix. If content is available at multiple PoPs, we choose the closest PoP based on hop count distance. We break ties randomly among PoPs with equal hop count distance.

We added a straightforward optimization to LRU where if a user terminates the request before 10% of the video (file) is viewed (downloaded), the content is not cached (and the rest of the file is not fetched); otherwise the entire file is downloaded and cached. This optimization is used since we observe in our traces that a user watching a video very often stops watching it after watching the initial period. A similar phenomenon is observed for large file downloads, but less frequently than video.

**OptR-LRU** uses a demand-oblivious placement, LRU, but it uses an demand-aware, optimized routing that is updated every three hours. The routing is computed by solving a multi-commodity flow problem identical to the traditional traffic engineering problem [12]. We assume that the NCDN measures the traffic matrix over the preceding three hours and computes routes that optimize the MLU for that matrix. The matrix incorporates the effect of the demand-oblivious placement and the implicit assumption is that the content demand and demand-oblivious placement result in a traffic matrix that does not change dramatically from one monitoring interval to the next—an assumption that also underlies traffic engineering as practiced by ISPs today.

**OptRP** computes a joint optimization of placement and routing once a day based on the previous day's content matrix using the MIP formulation of §3.2. **OptRP-Future** has oracular knowledge of the content matrix for the next day and uses it to calculate a joint optimization of placement and routing. OptRP and OptRP-Future are identical in all respects except that the former uses the content matrix of the past day while the latter has perfect future knowledge. These two schemes help us understand the value of future knowledge. In practice, it may be possible for an NCDN to obtain partial future knowledge placing it somewhere between the two extremes. For instance, an NCDN is likely to be informed beforehand of a major software release the next day (e.g., new version of the Windows) but may not be able to anticipate a viral video that suddenly gets "hot".

To determine the value of optimizing routing alone, we study the **InvCap-OptP-Future** scheme. This is a variant of OptRP-Future where InverseCap routing is used and content placement is optimized, rather than jointly optimizing both. This scheme is computed using the MIP formulation in §3.2 but with a simple modification (refer to [1] for the modified MIP formulation).

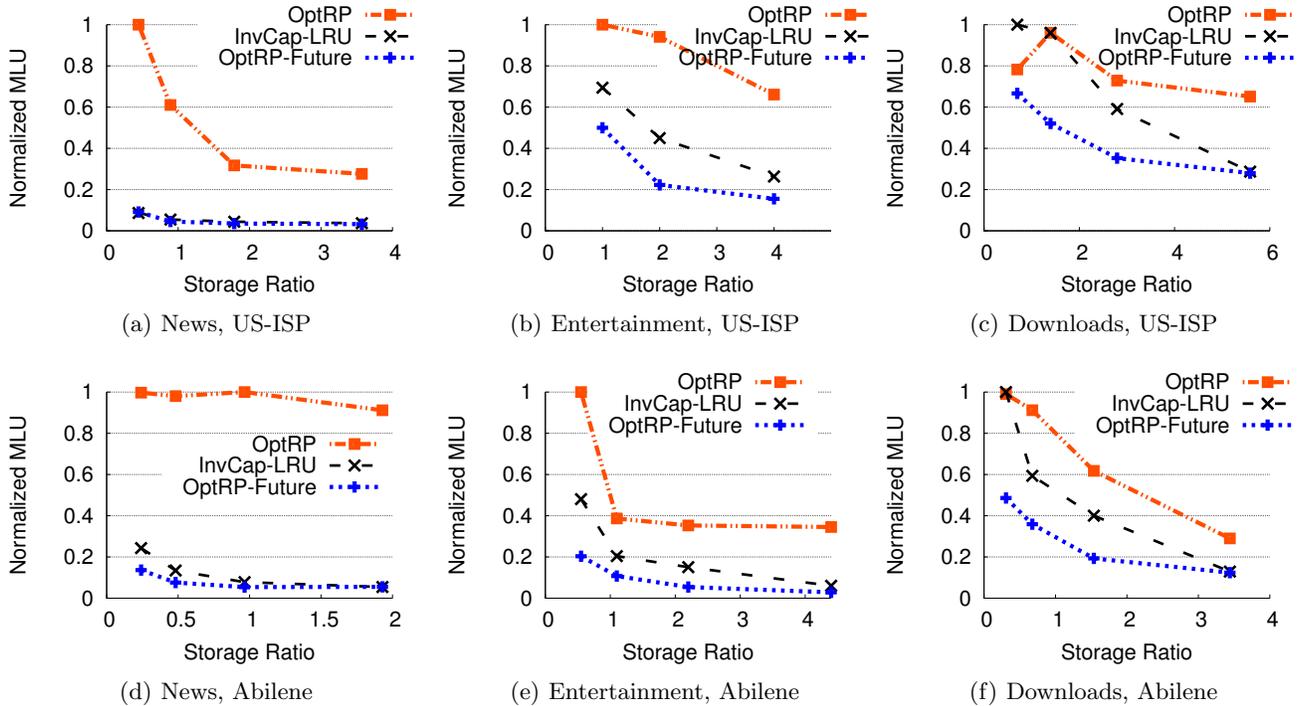

Figure 7: Demand-aware **OptRP** performs much worse than demand-oblivious **InvCap-LRU**. **OptRP-Future** performs moderately better than **InvCap-LRU** primarily at small storage ratios.

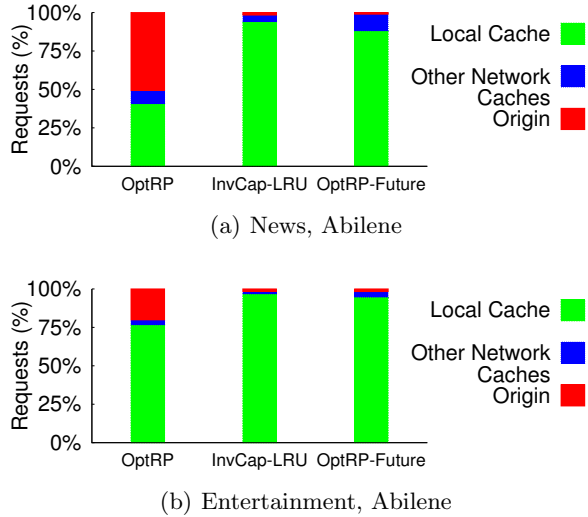

(a) News, Abilene

(b) Entertainment, Abilene

Figure 8: [Videos, Abilene] **OptRP** serves 50% and 21% of news and entertainment requests respectively from the origin. **InvCap-LRU** and **OptRP-Future** serve at most 2% from the origin.

For all schemes that generate a new placement each day, we implement the new placement during the low-traffic period from 4 AM to 7 AM EST. This ensures that the traffic generated due to changing the content placement occurs when the links are underutilized. For these schemes, the routing is updated each day at 7 AM EST once the placement update is finished.

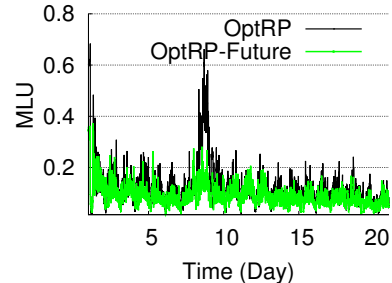

Figure 9: [Downloads, US-ISP] **OptRP** incurs a very high MLU on one "peak load" day.

### 5.3 Analysis of Video & Downloads Traffic

Figure 7 shows the results for the news, entertainment and downloads traces on Abilene and US-ISP. Our first observation is that a realistic demand-aware placement and routing scheme, **OptRP**, performs significantly worse than a completely demand-oblivious scheme, **InvCap-LRU**. **OptRP** has $2.2\times$ to $17\times$ higher MLU than **InvCap-LRU** even at the maximum storage ratio in each graph. **OptRP** has a high MLU because it optimizes routing and placement based on the previous day's content demand while a significant fraction of requests are for new content not accessed the previous day (see Figure 6). Due to new content, the incoming traffic from origin servers is significant, so the utilization of links near the exit nodes connecting to the origin servers is extremely high.

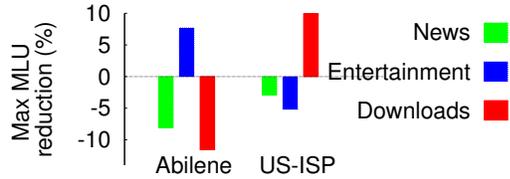

(a) MLU reduction with OptR-LRU compared to InvCap-LRU

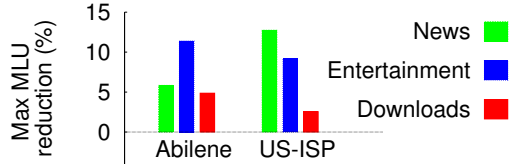

(b) MLU reduction with OptRP-Future compared to InvCap-OptP-Future

**Figure 10:** [All traces] Optimizing routing yields little improvement to MLU of either **InvCap-LRU** or **InvCap-OptP-Future**

The fraction of requests served from the origin is much higher for OptRP compared to InvCap-LRU and OptRP-Future on the news and the entertainment traces. Figure 8 shows that OptRP serves 50% and 21% of requests from the origin for news and entertainment respectively. In comparison, InvCap-LRU and OptRP-Future serve less than 2% of requests from the origin. Therefore, OptRP has a much higher MLU than both InvCap-LRU and OptRP-Future on the two traces.

The downloads trace differs from other traces in that, except for one day, the traffic is quite predictable based on the previous day's history. This is reflected in the performance of OptRP that performs nearly the same as OptRP-Future on all days except the ninth day of the trace (see Figure 9). The surge in MLU for OptRP on the ninth day is because nearly 20% of requests on this day is for new content consisting of highly popular software update releases (see Figure 6). The surge in MLU on this one day is mainly responsible for the poor performance of OptRP on the downloads trace.

Next, we observe that InvCap-LRU does underperform compared to OptRP-Future that has knowledge of future content demand. However, InvCap-LRU improves with respect to OptRP-Future as the storage ratio increases. The maximum difference between the two schemes is for the experiment with entertainment trace on US-ISP topology. In this case, at a storage ratio of 1, InvCap-LRU has twice the MLU of the OptRP-Future scheme; the difference reduces to 1.6× at a storage ratio of 4. This shows that when storage is scarce, demand-aware placement with future knowledge can significantly help by using knowledge of the global demand to maximize the utility of the storage. However, if storage is plentiful, the relative advantage of OptRP-Future is smaller.

An important implication of our results is that an NCDN should attempt to do demand-aware placement only if the future demand can be accurately known or estimated. Otherwise, a simpler demand-oblivious scheme such as LRU suffices.

How are the above conclusions impacted if InvCap-LRU were to optimize routing or OptRP-Future were to use InverseCap routing? To answer this question, we analyze the maximum reduction in MLU by using OptR-LRU over InvCap-LRU across all storage ratios in Figure 10. We similarly compare OptRP-Future and InvCap-OptP-Future. We find that OptR-LRU improves the MLU over InvCap-LRU by at most 10% across all traces suggesting that optimizing routing is of little value for a demand-oblivious placement scheme. OptRP-Future reduces the network cost by at most 13% compared to InvCap-OptP-Future. As we consider OptRP-Future to be the "ideal" scheme with full future knowledge, these results show that the best MLU can be achieved by optimizing content placement alone; optimizing routing adds little additional value.

Somewhat counterintuitively, the MLU sometimes increases with a higher storage ratio for the OptRP scheme. There are three reasons that explain this. First, the optimization formulation optimizes for the content matrix assuming that the demand is uniformly spread across the entire day, however the requests may actually arrive in a bursty manner. So it may be sub-optimal compared to a scheme that is explicitly optimized for a known sequence of requests. Second, the optimization formulation optimizes the MLU for the "smoothed" matrix, but the set of objects placed by the optimal strategy with more storage may not necessarily be a superset of the objects placed by the strategy with lesser storage at any given PoP. Third, and most importantly, the actual content matrix for the next day may differ significantly from that of the previous day. All of these reasons make the so-called "optimal" OptRP strategy suboptimal and in combination are responsible for the nonmonotonicity observed in the experiments.

### 5.4 Content Chunking

As chunking is widely used to improve content delivery [2, 24, 9], we repeated our experiments with smaller chunks of content. In these experiments, we split videos into chunks of 5 minute duration. The size of a video chunk depends on the video bitrate. For the downloads trace, we split content into chunks of size 50 MB.

Next, we describe our findings on the impact of chunking on NCDN strategies. We find that although chunking improves performance of both InvCap-LRU and OptRP-Future, it significantly improves the performance of InvCap-LRU relative to OptRP-Future (see Figure 11). Due to chunking, the maximum difference between the MLU of InvCap-LRU and OptRP-Future reduces from 2.5× to

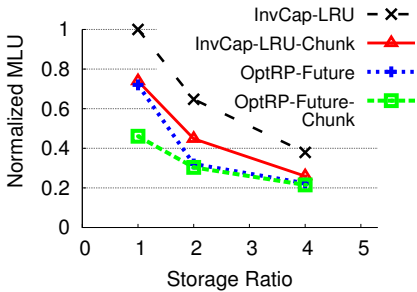 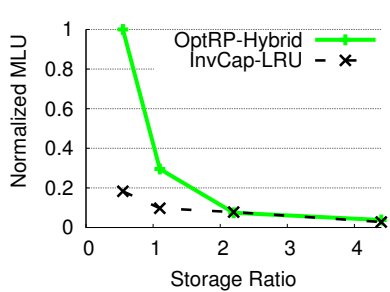 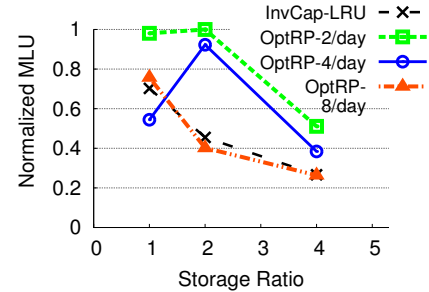

Figure 11: [Entertainment, US-ISP] Content chunking helps bridge the gap between **InvCap-LRU** and **OptRP-Future**.

Figure 12: [Entertainment, Abilene] Hybrid placement schemes perform at best as well as **InvCap-LRU**.

Figure 13: [Entertainment, US-ISP] **OptRP** does not outperform **InvCap-LRU** despite engineering 8 times a day.

1.4×. At the maximum storage ratio, InvCap-LRU is at most 20% worse compared to OptRP-Future. Our experiments on other traces and topologies (omitted for brevity) have qualitatively similar conclusions. An exception is the news trace, where chunking makes a small difference as more than 95% content is of duration less than our chunk size. Hence, chunking strengthens our conclusion that InvCap-LRU achieves close to the best possible network cost for an NCDN. Even with chunking, OptRP has up to 7× higher MLU compared to InvCap-LRU (not shown in Figure 11). This is because chunking does not help OptRP's primary problem of not being able to adapt effectively to new content, so it continues to incur a high cost.

### 5.5 Alternative Demand-aware Schemes

The experiments so far suggest that a demand-aware scheme that engineers placement and routing once a day based on the previous day's demand performs poorly compared to a demand-oblivious scheme, InvCap-LRU. Therefore, in this section, we evaluate the performance of two alternative demand-aware schemes.

First, we evaluate a hybrid placement scheme, which splits the storage at each node into two parts - one for a demand-aware placement based on the previous day's content demand (80% of storage) and the other for placing the content in a demand-oblivious LRU manner (20% of storage). This hybrid strategy is similar to that used in [3]. We find that InvCap-LRU performs either as well or better than the hybrid scheme. We also experimented with assigning a greater fraction of storage to demand-oblivious placement (omitted for brevity), but the above conclusions remain unchanged in those experiments. Of course, a carefully designed hybrid scheme by definition should perform at least as well as the demand-oblivious and demand-aware schemes, both of which are extreme cases of a hybrid strategy. However, we were unable to design simple hybrid strategies that consistently outperformed fully demand-oblivious placement and routing.

Next, we analyze the performance of demand-aware schemes that engineer placement and routing multiple times each day at equal intervals - twice/day, 4 times/day, and 8 times/day. In all cases, we engineer using the content demand in the past 24 hours. As Figure 13 shows, OptRP needs to engineer 8 times/day to match the performance of the InvCap-LRU scheme. In all other cases, InvCap-LRU performs better. In fact, the experiment shown here represents the best case for OptRP. Typically, OptRP performs worse even when engineering is done 8 times/day, e.g., on the news trace, we find OptRP incurs up to 4.5× higher MLU compared to InvCap-LRU even on engineering 8 times/day.

Executing a demand-aware placement requires considerable effort—measuring content matrix, solving a computationally intensive optimization, and moving content to new locations. Further, a demand-aware placement needs to be executed 8 times a day (or possibly more) even to match the cost achieved by a demand-oblivious strategy. Our position is that NCDNs are better served by opting for a much simpler demand-oblivious strategy and provisioning more storage, in which case, a demand-oblivious strategy already obtains a network cost close to the best a demand-aware strategy can possibly achieve.

### 5.6 Link-utilization-aware Redirection

This redirection scheme is designed to reduce link utilization in an NCDN. The idea is to periodically measure utilizations of all links and then redirect requests over the paths that have less utilized links. For example, if a PoP (say PoP A) can redirect a request to either of two PoPs, then we choose the PoP for which the utilization of the most utilized link along the path from that PoP to PoP A is the least. Link utilization levels that are necessary for this scheme can be monitored by an NCDN for its own network.

We experiment with two versions of the above algorithm, which differ in the frequency at which link utilization levels are measured across all PoPs. In the first version, InvCap-LRU-Current, all PoPs know the link utilizations of all links at every moment. In the second version, InvCap-LRU-30sec, link utilization is measured at all links every 30 seconds, and is updated across all PoPs. Our goal is to identify if link-utilization-aware redirection improves performance at all compared to the baseline; further optimizing the frequency of monitoring link utilization levels is beyond the scope of this paper.

Our experiments show that link-utilization-aware redirection does reduce the network cost compared to InvCap-LRU. InvCap-LRU-Current achieves between 10% to 21% lower MLU compared to InvCap-LRU (graph presented in [1]). In addition, InvCap-LRU-30sec matches the network cost of InvCap-LRU-Current on all traces except for the downloads trace on Abilene, in which case it still reduces the MLU between 7%-13% over InvCap-LRU.

Content chunking and link-utilization-aware redirection, in combination, nearly blur the difference between InvCap-LRU and OptRP-Future. After these optimizations are used, InvCap-LRU is at most 4% sub-optimal compared to OptRP-Future at the maximum storage ratio in each experiment. In some cases, InvCap-LRU even achieves a lower network cost than OptRP-Future, e.g, on the entertainment trace on the Abilene topology, InvCap-LRU performs 22% better than OptRP-Future.

## 5.7 Other Results and Implications

We summarize our findings from other experiments here and refer the reader to [1] for a full description.

**Request redirection to neighbors:** If each PoP redirects requests only to its one-hop neighbor PoPs before redirecting to the origin, InvCap-LRU incurs only a moderate (6%-27%) increase in the MLU. However, if a PoP redirects to no other PoPs but redirects only to the origin, the MLU for InvCap-LRU increases significantly (25%-100%). Thus, request redirection to other PoPs helps reduce network cost, but most of this reduction can be had by redirecting only to neighboring PoPs.

**Heterogenous storage:** Heterogenous storage at PoPs (storage proportional to the number of requests at a PoP in a trace, and other simple heuristics) increases the MLU compared to homogenous storage for both InvCap-LRU and OptRP-Future, and makes InvCap-LRU more sub-optimal compared to OptRP-Future. This leads us to conclude that our results above with homogeneous storage are more relevant to practical settings.

**OptR-LRU parameters:** Whether OptR-LRU updates routing every 3 (default), 6, or 24 hours, makes little difference to its performance. Further, whether OptR-LRU optimizes routing using traffic matrix measured over the immediately preceding three hours (default) or using traffic matrices measured the previous day, its network cost remains nearly unchanged. These experiments reinforce our finding that optimizing routing gives minimal improvement over InvCap-LRU.

**Number of exit nodes:** When the number of network exit nodes is increased to five or decreased to one, our findings in §5.3 remain qualitatively unchanged.

**Link failures:** The worst-case network cost across all single link failures for InvCap-LRU as well as OptRP-Future is approximately twice compared to their network costs during a failure-free scenario. Comparing the failure-free scenario and link failure scenarios, the relative sub-optimality of InvCap-LRU with respect to OptRP-Future remains the same at small storage ratios but reduces at higher storage ratios.

**A note on the need for traffic engineering:** Overall, our results suggest that optimizing routing yields little improvement to network cost for the NCDN portion of the traffic, but this finding does not imply that traffic engineering by ISPs is unnecessary. ISPs route transit traffic in addition to NCDN traffic, and they do not control either content placement or request redirection for transit traffic, therefore, traditional traffic engineering methods, e.g., OSPF traffic engineering, can reduce the network cost due to transit traffic. The benefit of traffic engineering, or lack thereof, depends on the fraction of transit traffic vs. NCDN traffic in an ISP.

## 6. RELATED WORK

Traffic engineering and content distribution have both seen an enormous body of work over more than a decade. To our knowledge, our work is the first to pose the NCDN problem, wherein a single entity seeks to address both concerns, and empirically evaluate different content-aware traffic engineering strategies.

**Joint optimization:** Recent work has explored the joint optimization of traffic engineering and "content distribution", where the latter term refers to the *server selection* problem. P4P (Xie et al. [30]) shows that P2P applications can improve their performance and ISPs can reduce the MLU and interdomain costs, if P2P applications adapt their behavior based on hints supplied by ISPs. Jiang et al. [18] and DiPalantino et al. [10] both study the value of joint optimization of traffic engineering and content distribution versus independent optimization of each. CaTE (Frank at al. [14]), like P4P, shows that a joint optimization can help both ISPs and content providers improve their performance. These works equate content distribution to server selection (or request redirection in our parlance), while the NCDN problem additionally considers content placement itself as a degree of freedom. As our results show, optimizing placement is powerful and can single-handedly reduce the MLU significantly even in conjunction with simple request redirection and routing strategies.

**Placement optimization:** Applegate et al. [3] study the content placement problem for a VoD system that seeks to minimize the aggregate network bandwidth consumed. However, they assume a *fixed routing* in the network, while one of our contributions is to assess the relative importance of optimizing routing and optimiz-ing placement in an NCDN.

Furthermore, they find that an optimized, demand-aware placement with a small local cache (similar to our "hybrid" strategy in §5.5) outperforms LRU. In contrast, our experiments suggest otherwise. There are three explanations for this disparity. First, their workload seems to be predictable even at weekly time scales, whereas the Akamai CDN traces that we use show significant daily churn. Second, their scheme has some benefit of future knowledge and is hence somewhat comparable to our OptRP-Future. For a large NCDN, obtaining knowledge about future demand may not be practical for all types of content, e.g., breakout news videos. Finally, our analysis suggests that LRU performs sub-optimally only at small storage ratios, and the difference between LRU and OptRP-Future reduces considerably at higher storage ratios (not considered in [3]).

**Traffic engineering:** Several classes of traffic engineering schemes such as OSPF link-weight optimization [12], MPLS flow splitting [11], optimizing routing for multiple traffic matrices [29, 31], online engineering [19, 11], and oblivious routing [4, 6], have been studied. All of these schemes assume that the demand traffic is a given to which routing must adapt. However, we find that an NCDN is in a powerful position to change the demand traffic matrix, so much so that even a naive scheme like InverseCap, i.e., no engineering at all, suffices in conjunction with a judicious placement strategy and optimizing routing further adds little value. In this respect, our findings are comparable in spirit to Sharma et al. [25]. However, they focus on the impact of location diversity, and show that even a small, fixed number of randomly placed replicas of each content suffice to blur differences between different engineering schemes with respect to a capacity metric (incomparable to MLU), but find that engineering schemes still outperform InverseCap routing.

## 7. CONCLUSIONS

We posed and studied the content-aware traffic engineering problem where content distribution and traffic engineering decisions are optimized jointly as opposed to being optimized separately as is done today. This paradigm is of key importance to NCDNs who own and manage the infrastructure for content distribution as well as the underlying network. Our trace-driven experiments using extensive access logs from the world's largest CDN and real ISP topologies resulted in the following key conclusions. First, simple demand-oblivious schemes for routing and placement, such as Inverse-Cap and LRU, outperformed sophisticated placement and routing schemes computed once-a-day based on the prior day's demand. Second, traffic demand can be "shaped" by effective content placement to the extent that the effectiveness of traditional traffic engineering is marginalized. Finally, we studied the value of the future knowledge of demand for placement and routing decisions. While future knowledge helps, what is perhaps surprising is that a small amount of additional storage and simple optimizations such as content chunking and link-utilization-aware redirection allows simple, demand-oblivious schemes to perform as well as demand-aware ones with future knowledge.

## 8. ACKNOWLEDGMENTS

The authors would like to thank S. Shunmuga Krishnan, Yin Lin, and Bruce M. Maggs for their help with the Akamai trace data collection.